\documentclass[preprint,12pt]{elsarticle}

\usepackage{graphicx}
\usepackage{latexsym}
\usepackage{stmaryrd} 
\usepackage{amsmath}
\usepackage{amssymb} 

\usepackage{amssymb}

\begin{document}
\newcommand{\eq}{\begin{equation}}
\newcommand{\eqe}{\end{equation}}

\begin{frontmatter}






\title{ Self-similar analytic solution of the two dimensional
Navier-Stokes equation with a non-Newtonian type of viscosity }

\author[label1]{Imre Ferenc Barna} and 
\author[label2]{Gabriella Bogn\'ar}
\address[label1]{ Wigner Research
Center of the Hungarian Academy of Sciences
 Konkoly-Thege \'ut 29 - 33, 1121 Budapest, Hungary}
\address[label2]{University of Miskolc, Miskolc-Egyetemv\'aros 3515, Hungary}
\begin{abstract}
We investigate the two dimensional incompressible Navier-Stokes(NS) and the
continuity equations in Cartesian coordinates and Eulerian description for non-Newtonian fluids.
As a non-Newtonian viscosity we consider the  Ladyzenskaya model with a non-linear velocity dependent stress tensor. The key idea is the multi-dimensional generalization of the well-known
self-similar Ansatz, which was already used for non-compressible and compressible viscous
flow studies. The geometrical interpretations of the trial function is also discussed.
We compare our recent results to the former Newtonian ones.  
\end{abstract}
\end{frontmatter}
\section{Introduction}
Dynamical analysis  of viscous fluids is a never-ending crucial problem.  A large part of real fluids do not strictly follow Newtons's law and are aptly called non-Newtonian fluids. In most cases, fluids can be  described by more complicated governing rules, which means that the viscosity has some additional density, velocity or temperature dependence or even all of them.
General introductions to the physics of  non-Newtonian fluid can be found in \cite{giova,bohme}.
In the following, we will examine the properties of a Ladyzenskaya type non-Newtonian fluid \cite{ladyz,bae}.
Additional temperature or density dependent viscosities will not be considered in the recent study.
There are some analytical studies available for non-Newtonian flows in connections with the boundary layer theory, which shows some similarity to our recent problem \cite{ishak,benlas}.
The heat transfer in the  boundary layer of a non-Newtonian Ostwald-de Waele power law fluid was investigated with self-similar Ansatz from one of us in details  as well \cite{gabi1}.
We use the two-dimensional generalization of the well-known self-similar Ansatz \cite{sedov,barenb,zeld}, which was already used to investigate the three dimensional  the non-compressible Newtonian NS \cite{barna1} and the compressible Newtonian NS
\cite{barna2} equations.   We compare our results to the former Newtonian cases.
In the last part of our manuscript we present a series solution for the ordinary differential equation(ODE), which was obtained with the help of the self-similar Ansatz. 

\section{Theoretical background}
The Ladyzenskaya \cite{ladyz} model of non-Newtonian fluid dynamics can be formulated in the most general  vectorial form of
\begin{eqnarray}
\rho \frac{\partial {\bf{ u}}_i}{\partial t} &+& \rho {\bf{u}}_j \frac{\partial {\bf{u}}_i}{\partial x_j} = - \frac{p}{\partial{ \bf{x}}_i}
+ \frac{\partial  \Gamma_{ij}}{\partial{\bf{ x}}_j}  + \rho{\bf{ a}}_i \nonumber  \\
\frac{\partial{\bf{ u}}_j}{\partial{\bf{ x}}_j} &= & 0  \nonumber  \\
\Gamma_{ij}   &  \overset{Def} {=}& (\mu_0 +  \mu_1  |E(\nabla{\bf{ u}})|^r ) E_{ij}(\nabla{\bf{ u}}) \nonumber \\
E_{ij}(\nabla {\bf{u}})&  \overset{Def}{=}& \frac{1}{2} \left (  \frac{\partial {\bf{u}}_i  }{\partial {\bf{x}}_j}+    \frac{\partial {\bf{u}}_j }{\partial {\bf{x}}_i} \right)
\end{eqnarray}
where $ \rho, {\bf{u}}_i,p,{\bf{a}}_i, \mu_0, \mu_1,r$ are the density, the two dimensional velocity field, the pressure, the external force, the dynamical viscosity, the consistency index and the flow behavior index. The last one is a dimensionless parameter of the flow.
(To avoid further misunderstanding, we use ${\bf{a}}$ for external field instead of the letter $g$, which is reserved for a self-similar solution.)
The $E_{ij}$ is the Newtonian linear stress tensor, where {\bf{x}}(x,y) are the Cartesian coordinates.
The Einstein summation is used for the j subscript.
In our next model, the exponent  should be $r>-1$.
This general description incorporates the following five different fluid models:
 \begin{eqnarray}
\mbox{Newtonian for}    & \mu_0 > 0, \mu_1 = 0,  \nonumber  \\
\mbox{Rabinowitsch for} &    \mu_0 , \mu_1 > 0,  r = 2, \nonumber  \\
\mbox{Ellis for}   & \mu_0 , \mu_1 > 0,  r>0,  \nonumber  \\
\mbox{Ostwald-de Waele for } &  \hspace*{1cm}   \mu_0 =0, \mu_1 > 0,  r>-1,  \nonumber  \\
\mbox{Bingham for}   & \mu_0, \mu_1 > 0,  r=-1.
\end{eqnarray}
For $\mu_0=0$ if $r<0$ then it is called a pseudo-plastic fluid, if $r>0$ it is a dilatant fluid \cite{bohme}. \\ 
In pseudoplastic or shear thinning fluid the apparent viscosity decreases with increased stress.
    Examples are: nail polish, whipped cream, ketchup, molasses, syrups, latex paint, ice, blood, some silicone oils, some silicone coatings and
paper pulp in water. For the paper pulp the numerical  Ostwald-de Waele parameters are
$\mu_1 = 0.418, r = -0.425$ \cite{bohme}. 
Parameters of a film foam with carbon dioxides are  $\mu_1 = 0.15, r = -0.52$ \cite{du}.  \\ 
In dilatant or shear thickening fluid, the apparent viscosity increases with increased stress.
Typical examples are suspensions of corn starch in water (sometimes called oobleck) or sand in water.  \\
We set the external force to zero in our investigation ${\bf{a_i}} = 0$.  In two dimensions, the absolute value of the stress tensor reads:
\eq
|E| = [u_x^2 + v_y^2 +1/2(u_y + v_x)^2 ]^{1/2},
\eqe
where the ${\bf{u}}(u,v)$  coordinate notation is used from now on.
For a better transparency - instead of the  the usual partial derivation notation -
we apply subscripts for partial derivations.
(Note, that in three dimensions the absolute value of the stress tensor would be much complicated containing six terms instead of three.)
Introducing the following compact notation
\eq
L = \mu_0 + \mu_1 |E|^r,
\eqe
our complete two dimensional NS system for incompressible fluids can be formulated much clearer:
 \begin{eqnarray}
u_x  + v_y = 0,  \nonumber \\
u_t  + uu_x + vu_y = -p_x/\rho  +  L_x u_x + Lu_{xx}  + \frac{L_y}{2}
(u_y + v_x) + \frac{L}{2}(u_{yy}+v_{xy}), \nonumber \\
v_t  + uv_x + vv_y = -p_y/\rho  +  L_y v_y + Lv_{yy}  + \frac{L_x}{2}
(u_y + v_x) + \frac{L}{2}(v_{xx}+u_{xy}),
\label{ezkell}
\end{eqnarray}
which is our starting point for the next investigation.  \\
We apply the  physically relevant self-similar Ansatz to system (\ref{ezkell}).
The form of the one-dimensional self-similar Ansatz is given in \cite{sedov,barenb,zeld}
\eq
T(x,t)=t^{-\alpha}f\left(\frac{x}{t^\beta}\right):=t^{-\alpha}f(\omega),
\label{self}
\eqe
where $T(x,t)$ can be an arbitrary variable of a partial differential equation(PDE) and $t$ means time and $x$ means spatial
dependence.
The similarity exponents $\alpha$ and $\beta$ are of primary physical importance since $\alpha$
represents the rate of decay of the magnitude $T(x,t)$, while $\beta$  is the rate of spread
(or contraction if  $\beta<0$ ) of the space distribution for $t > 0 $.
The most powerful result of this Ansatz is the fundamental or
Gaussian solution of the Fourier heat conduction equation (or for Fick's
diffusion equation) with $\alpha =\beta = 1/2$.
This transformation
is based on the assumption that a self-similar solution
exists, i.e., every physical parameter preserves its
shape during the expansion. Self-similar solutions usually
describe the asymptotic behavior of an unbounded or a far-field
problem; the time $t$ and the space coordinate $ x$ appear
only in the combination of  $f(x/t^{\beta})$. It means that the existence
of self-similar variables implies the lack of characteristic
length and time scales. These solutions are usually not unique and
do not take into account the initial stage of the physical expansion process.
These kind of solutions  describe the intermediate asymptotic of a problem: they hold when the precise initial
conditions are no longer important, but before the system has reached its final steady state. They are much simpler than
the full solutions and so easier to understand and study in different regions of parameter space. A final reason for studying
them is that they are solutions of a system of ODE and hence do not suffer the extra inherent numerical
problems of the full partial differential equations. In some cases self-similar solutions helps to understand diffusion-like properties
or the existence of compact supports of the solution.  \\ 
At first we introduce the two dimensional generalization of the self-similar Ansatz,
(\ref{self}) which might have the form of 
\eq
T(x,z,t) = t^{-\alpha} f\left(  \frac{F(x,y)}{t^{\beta}} \right)
	\eqe 
where $F(x,y)$ could be understood as an implicit parametrization of a one-dimensional space curve with continuous first and second derivatives. 
 In our former studies \cite{barna1,barna2}, we explain that for the non-linearity in the Navier-Stokes equation unfortunately only the $ F(x,z ) = x+y+c$ function is valid in Cartesian coordinates. 
It basically comes from the symmetry properties of the system like $u_x = u_y$. 
(Other locally orthogonal coordinate systems have not been investigated yet.) 

Every two dimensional flow problem can be reformulated with the help of the stream function $\Psi$ via $u = \Psi_y$ and $v = -\Psi_x$, 
 which automatically fulfills the continuity equation.
The system of  (\ref{ezkell}) is now reduced to the following two PDEs
 \begin{eqnarray}
\Psi_{yt} + \Psi_y\Psi_{yx} - \Psi_x\Psi_{yy} =   - \frac{p_x}{\rho} + (L\Psi_{yx})_x + \left[ \frac{L}{2} (\Psi_{yy} - \Psi_{xx}) \right]_y  \nonumber  \\
-\Psi_{xt} - \Psi_y\Psi_{xx} + \Psi_x\Psi_{xy} =   - \frac{p_y}{\rho} +      \left[ \frac{L}{2} (\Psi_{yy} - \Psi_{xx}) \right]_x -  (L\Psi_{yx})_y
\label{stream}
\end{eqnarray}
with $L = \mu_0 + \mu_1 \left[  2 \Psi_{xy}^2 +   \frac{1}{2} (\Psi_{yy}-\Psi_{xx})^2 \right]^{r/2}.$  

Now, search the solution of this PDE system such as:  $\Psi = t^{-\alpha}f(\eta),    \hspace*{1mm} p = t^{-\epsilon}h(\eta), \hspace*{1mm} \eta = \frac{x+y}{t^{\beta}}, $ 
where all the exponents $\alpha,\beta,\gamma$ are real numbers. (Solutions with
integer exponents are called self-similar solutions of the first kind and they can be obtained from dimensional argumentation as well.)

Unfortunately, the constraints, which should fix the values of the exponents become contradictory, therefore no unambiguous ODE can be formulated.
This means that the PDE and the stream function does not have self-similar solutions. In other words the stream function has no diffusive property.
This is a very instructive example of the applicability of the trial function of (\ref{self}). 
 
Let's return to the original system  of (\ref{ezkell}) and  try the Ansatz of 
\eq
u = t^{-\alpha}f(\eta), \hspace*{1cm} v = t^{-\delta}g(\eta),\hspace*{1cm}  p = t^{-\epsilon}h(\eta), \hspace*{1cm} \eta = \frac{x+y}{t^{\beta}}
\eqe
where all the exponents are real number again and  $f,g,h$ are called the shape functions. 
The next step is to determine the exponents. 
From the continuity equation we simple get arbitrary $\beta$ and $\delta = \alpha$ relations.
The two  NS dictate additional constraints.
(We skip the trivial case of $ \mu_0 \ne 0, \mu_1 = 0$, which was examined in our former paper as the Newtonian fluid.  \cite{barna1})
Finally, we get
\begin{equation}
 \mu_0 = 0, \>\>  \mu_1 \ne 0, \>\> \alpha = \delta = (1+r)/2, \>\> \beta = (1-r)/2, \>\> \epsilon = r+1.
\label{exp}
\end{equation}
Note, that $r$ remains free, which describes various fluids with diverse physical properties, this meets our expectations.
For the Newtonian NS equation, there is no such free parameter and in our former investigation we got fixed exponents with a value of  $1/2$.
For a physically relevant solutions, which is spreading and decaying in time  all the exponents  Eq. (\ref{exp}) have to be positive determining the $ -1 < r <1 $ range.
This can be understood as a kind of restricted Ostwald-de Waele-type fluid.
After some algebra a  second order non-autonomous non-linear ODE remains
\eq
  \mu_1(1 + r) f''[2f']^{r/2} + \frac{(1-r)}{2}\eta f' + \frac{(1+r)}{2}f =0,
\label{vegs}
\eqe
where prime means derivation with respect to $\eta$.
Note that for the numerical value $ r=0$ we get back the ODE of the Newtonian NS equation for two dimensions.
In three dimensions the ODE reads:
\eq
9\mu_0 f'' - 3(\eta +c)f' + \frac{3}{2}f(\eta) - \frac{c}{2} +a = 0.
\label{kummerdif}
\eqe
Its solutions are the Kummer functions \cite{abr}
 \begin{eqnarray}
f = c_1 \cdot KummerU \left(-\frac{1}{4},\frac{1}{2},\frac{(\eta+c)^2}{6 \mu_0} \right)  +  \nonumber \\ 
 c_2 \cdot KummerM \left( -\frac{1}{4},\frac{1}{2},\frac{(\eta+c)^2}{6
\mu_0} \right) + \frac{c}{3} -\frac{2a}{3},
\label{kum}
 \end{eqnarray}
where $c_1$ and $c_2$ are integration constants, $c$ is the  mass flow rate, and $a$ is the  external field.
These functions have no compact support.  The corresponding velocity component however, decays for large time like $ v \sim 1/t $ for $ t  \rightarrow  \infty $, which makes these results  physically reasonable.
A detailed analysis of (\ref{kum}) was presented in \cite{barna1}. A similar investigation was performed for compressible Newtonian fluids in \cite{barna2}, where the results are given by the Whitakker functions, which have strong connections to the Kummer functions as well. 

Unfortunately, we found no analytic solution or integrating factor for (\ref{vegs}) at any arbitrary values of $r$.
For sake of completeness, we mention that with the application of the symmetry properties of 
the Ansatz $u_{xx }=u_{xy}=u_{yy}=-v_{xx }=-v_{xy}=-v_{yy} $ 
the following closed form can be derived for the pressure field 
\eq
h = \rho (\mu_1 2^{r/2 +1}f'^{r+1} + f\eta  -\tilde{c}_1f    ) + \tilde{c}_2,  
\eqe
where $\tilde{c}_1$ and $\tilde{c}_2$ are the usual integration constants.  
\section{Phase Plane Analysis}
Applying the transition theorem, a second order ODE is always
equivalent to a first order ODE system. Let us substitute $f' =l,
f'' = l' $, then
\begin{eqnarray}
  f'& =& l,               \nonumber  \\
 l' &=&  -\left( \frac{(1-r)}{2}\eta l + \frac{(1+r)}{2}f \right) /  \left( \mu_1(1 + r)  2^{r/2} l^r \right),
\label{sys}
\end{eqnarray}
where prime still means derivation with respect to $\eta$.
This ODE system is still non-autonomous and there is no general theory to investigate such phase portraits.
We can divide the second equation of (\ref{sys}) by the first one  to get a new  ODE, where the former independent variable $\eta$ becomes a free real parameter
\eq
\frac{dl}{df} =  -\left( \frac{(1-r)}{2}\eta l + \frac{(1+r)}{2}f \right) /  \left( \mu_1(1 + r)  2^{r/2} l^{r+1} \right).
\label{diagr}
\eqe
Figure 1 shows the phase portrait diagram of  (\ref{diagr}) for  water pulp the material parameters are $ r = -0.425$ and  $\mu_1 = 0.18$.
 We consider $\eta = 0.03$  as the "general time variable" to be positive as well.
With the knowledge of  the exponent range $ -1 < r < +1$, two general properties of the phase space can be understood
by analyzing  Eq.  (\ref{diagr}): \\
Firstly, the derivative $df /dl$ is zero at zero nominator values, which means \\
 $l = -(1+r)/(1-r) f/\eta$. This one is a straight line passing through the origin
with gradient of  $0 < (1+r)/(1-r) < \infty $  for $ -1 < r < +1$. On Figure 1, the numerical value of the gradient is $ -0.403/\eta = -13.5$.   \\
Secondly,  the derivative  $df/dl$ or the direction field is not defined for any negative $l$ values because the power function $l^{-(r+1)}$ in the denominator is not defined for negative $l$ arguments. We cannot extract a non-integer root from a negative number.
The denominator is always positive.  \\
These properties   explain  that there are two kinds of possible trajectories or solutions exist in the phase space.
One type has a compact support, and the other has a finite range.
We may consider the  $x$ axis as the velocity $ f \sim v(\eta)$ and $y$ axis as the $ f' \sim  a(\eta) $ acceleration for a fixed scaled time $\eta = const.$
It also means that the possible velocity and accelerations for a general time cannot be independent from each other. 
The factors of the second derivative  $f''$  show some similarity to the porous media equation, where the diffusion coefficient has also an exponent.  This is the essential original responsibility
for the solution with compact support \cite{robi}. This is our main result. 
\begin{figure}
\hspace*{0.4cm}
\scalebox{0.6}{
\rotatebox{0}{\includegraphics{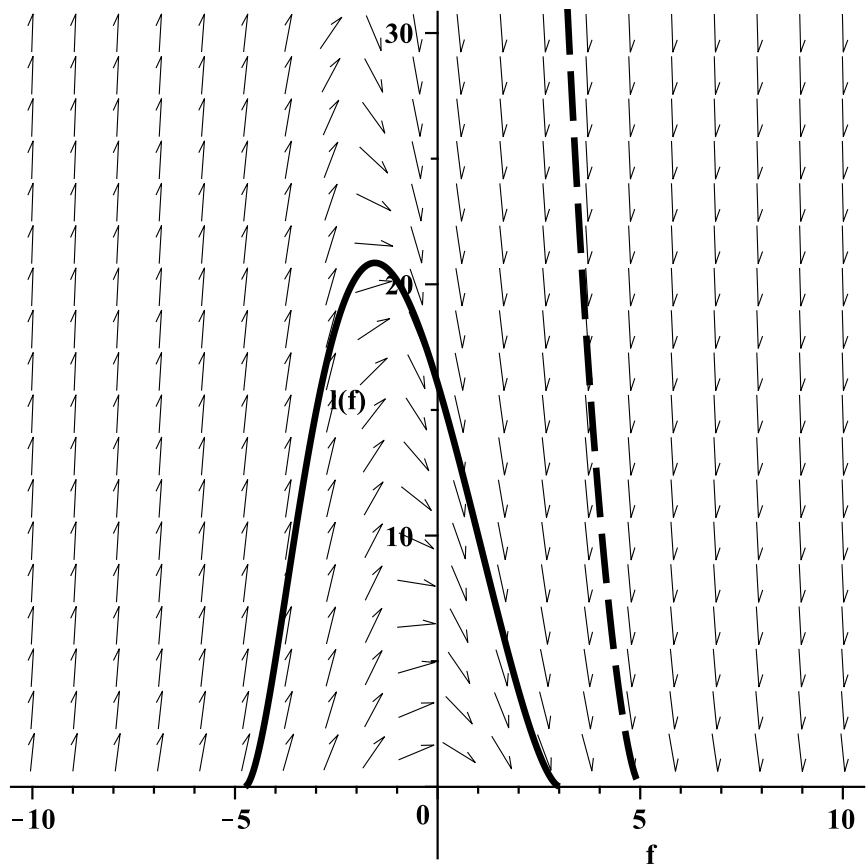}}}
\vspace*{-0.4cm}
\caption{ The phase portrait diagram of  (\ref{diagr})   for $\eta = 0.03, r = -0.425$ and  $\mu_1 = 0.18$. The two different kind of trajectories: one with compact support (solid line), and the  other one with compact range (dashed line).}
\label{kettes}       
\end{figure}

\section{Approximate Solutions}
Our objective is to show the existence of analytic solutions to the
differential equation (\ref{vegs}) and to determine the approximate
local solution $f(\eta )$. We use the shooting method and give the
conditions  at $\eta =0$  with initial conditions
\begin{equation}
f(0)=A, {\ \ \ \ }f^{\prime }(0)=B. \label{13}
\end{equation}%
We will consider (\ref{vegs}) as a system of certain differential
equations, namely, the special Briot-Bouquet differential equations.
For this type of differential equations we refer to the book by E.
Hille \cite{hille}. In order to establish the existence of a power
series representation of $f(\eta )$ about $\eta =0$ we apply the
following theorem: \\
Briot-Bouquet Theorem \cite{briot}: Let us assume that for the system
of equations
\begin{equation}
\left.
\begin{array}{c}
\xi \frac{dz_{1}}{d\xi }=u_{1}(\xi ,z_{1}(\xi ),z_{2}(\xi )), \\
\xi \frac{dz_{2}}{d\xi }=u_{2}(\xi ,z_{1}(\xi ),z_{2}(\xi )),%
\end{array}%
\right\}   \label{14}
\end{equation}%
where functions $u_{1}$ and $u_{2}$ are holomorphic functions of $\xi ,$ $%
z_{1}(\xi ),$ and $z_{2}(\xi )$ near the origin, moreover
\begin{equation}
u_{1}(0,0,0)=u_{2}(0,0,0)=0,
\end{equation}%
then a holomorphic solution of (\ref{14}) satisfying the initial conditions $%
z_{1}(0)=0,$ $z_{2}(0)=0$ exists if none of the eigenvalues of the
matrix
\begin{equation}
\left[
\begin{array}{cc}
\left. \frac{\partial u_{1}}{\partial z_{1}}\right\vert _{(0,0,0)} &
\left.
\frac{\partial u_{1}}{\partial z_{2}}\right\vert _{(0,0,0)} \\
\left. \frac{\partial u_{2}}{\partial z_{1}}\right\vert _{(0,0,0)} &
\left.
\frac{\partial u_{2}}{\partial z_{2}}\right\vert _{(0,0,0)}%
\end{array}%
\right]  \label{M}
\end{equation}%
is a positive integer. \\  
Briot-Bouquet theorem ensures the existence of formal solutions
\begin{equation}
z_{1}={\sum }_{k=0}^{\infty } a_{k}\xi ^{k}, { \ \ \ \ }%
z_{2}={\sum }_{k=0}^{\infty } b_{k}\xi ^{k}
\end{equation}%
for system (\ref{14}), and also the convergence of formal solutions.
\newline This theorem has been successfully applied to the
determination of local
analytic solutions of different nonlinear initial value problems \cite{[4]}-%
\cite{[6]}. \newline Let us consider the initial value problem
(\ref{vegs}), (\ref{13}) and take its solution in the form
\begin{equation}
f(\eta )=\eta ^{\alpha }Q\left( \eta ^{\beta }\right) , {\ }\eta \in
\left( 0,\;\eta _{c}\right) ,  \label{15}
\end{equation}%
where function
\begin{equation}
Q\in C^{2}(0,\;\eta _{c})
\end{equation}%
for some positive value $\eta _{c}$. Substituting
\begin{equation}
f(\eta )=\eta ^{\alpha }Q\left( \eta ^{\beta }\right)
\end{equation}%
into (\ref{vegs}) one can get
\begin{eqnarray}
K\frac{1}{\eta }\left\{ \beta \left[ \left( \beta +1\right) \eta
^{\beta }Q^{\prime }+\beta \eta ^{2\beta }Q^{\prime \prime }\right]
\right\} \left( Q+\beta \eta ^{\beta }Q^{\prime }\right)
^{\frac{r}{2}}+ \nonumber \\ \frac{1-r}{2}\eta \left( Q+\beta \eta ^{\beta
}Q^{\prime }\right) +  
\frac{1+r}{2}\eta Q=0 \label{16}
\end{eqnarray}%
for $\alpha =1$. Let us introduce the new variable $\xi $ such as
\begin{equation}
\xi =\eta ^{\beta }
\end{equation}%
and function $Q$ as follows
\begin{equation}
Q(\xi )=a_{0}+a_{1}\xi +z(\xi ),
\end{equation}%
where $a_{0},a_{1}$ are real constants, and
\begin{equation}
z\in C^{2}(0,\eta _{c}^{\beta }),
\end{equation}%
$z(0)=0,$ $z^{\prime }(0)=0$. We make difference between two cases:
$A=0$ and $A\neq 0$. \\ 
If $A=0$ then $Q$ fulfills the following properties $Q(0)=a_{0},$
$Q^{\prime }(0)=a_{1},$  $Q^{\prime \prime }(\xi )=z^{\prime \prime
 }(\xi ).$ From the initial condition $f^{\prime
}(0)= Q(0)$ and we have $ a_{0}=B.$ Applying the Briot-Bouquet
theorem,  ({\ref{16}})  yields
\begin{equation}
\xi z^{\prime \prime }\left( \xi \right) =-\frac{1}{\beta }\left[
\left(
\beta +1\right) Q^{\prime }+\xi ^{\frac{2}{\beta }-1}\frac{Q+\frac{1-r}{2}%
\beta \xi Q^{\prime }}{K\beta \left( Q+\beta \xi Q^{\prime }\right) ^{\frac{r%
}{2}}}\right],
\end{equation}
with $K=\mu _{1}(1+r)2^{\frac{r}{2}}$. Therefore, $\beta =2$. We
restate the second order differential equation in (\ref{16}) as a
system of two equations
\begin{equation}
\left.
\begin{array}{c}
u_{1}(\xi ,z_{1}(\xi ),z_{2}(\xi ))=\xi \,z_{1}^{\prime }(\xi )
\\
u_{2}(\xi ,z_{1}(\xi ),z_{2}(\xi ))=\xi \,z_{2}^{\prime }(\xi )
\end{array}%
\right\}
\end{equation}%
with choosing
\begin{equation}
\left.
\begin{array}{c}
z_{1}(\xi )=z(\xi ) \\
z_{2}(\xi )=z^{\prime }(\xi ) %
\end{array}%
\right\} \ {\ }\ {\ with}\;\,\ {\ \ }\left.
\begin{array}{c}
z_{1}\left( 0\right) =0 \\
z_{2}\left( 0\right) =0 %
\end{array}%
\right\}
\end{equation}%
and
\begin{equation}
u_{1}(\xi ,z_{1}(\xi ),z_{2}(\xi ))=\xi \,z_{2}, \label{18}
\end{equation}%
\begin{eqnarray}
u_{2}(\xi ,z_{1}(\xi ),z_{2}(\xi ))=\xi \,z_{2}^{\prime }
=-\frac{1}{2}\left[ 3\left( a_{1}+z_{2}\right) +   \right] - 
 \nonumber \\    \left[ \frac{%
a_{0}+(2-r)a_{1}\xi +z_{1}+(1-r)\xi z_{2}}{2K\left( a_{0}+3a_{1}\xi
+z_{1}+2\xi z_{2}\right) ^{\frac{r}{2}}}\right].
\end{eqnarray}%
In order to satisfy the conditions
\begin{equation}
u_{1}(0,0,0,0)=u_{2}(0,0,0,0)=0
\end{equation}%
in the Briot-Boquet theorem the following connection yields
\begin{equation}
a_{1}=-\frac{a_{0}^{1-\frac{r}{2}}}{6K}.
\end{equation}%

Therefore, the eigenvalues of matrix (\ref{M}) at $(0,0,0)$ are $0$.
Since all  eigenvalues are non-positive, referring to the
Briot-Bouquet theorem we obtain the existence of unique analytic
solutions $z_{1}$ and $z_{2}$  near zero. Thus, there exists a
formal solution
\begin{equation}
f(\eta )={\eta }\sum\limits_{k=0}^{\infty }a_{k}\eta ^{2k},
\label{25}
\end{equation}%
where the first two coefficients are already known. \newline For the
determination of coefficients $a_{k},\;\,k>2$, we shall use the
J.C.P. Miller formula \cite{miller}:
\begin{equation}
\left[ \sum\limits_{k=0}^{L}c_{k}x^{k}\right] ^{r/2}=\sum%
\limits_{k=0}^{\frac {r}{2}L}d_{k}(r)x^{k},
\end{equation}%
where $d_{0}(r)=1\,$ for $c_{0}=1,$ and
\begin{equation}
d_{k}(r)=\frac{1}{k}\,\sum\limits_{j=0}^{k-1}[\frac {r}{2}(k-j)-j]d_{j}(r)c_{k-j},%
{\ \ \ \ }(k\geq 1).  \label{26}
\end{equation}%
Applying (\ref{26}) we get recursion formula for the determination
of $a_{k}$. Comparing the proper coefficients of $\eta $, one can
have the  values of $a_{k}$ for some  $k$; e.g.,
\begin{equation}
a_{2}=\frac{2-r}{120K^{2}}a_{0}^{1-r}-\frac{1}{80K^{2}}a_{0}^{2-r}.
\end{equation}
The coefficients are obtained by this method for the power series
approximation as
\begin{equation}
f\left( \eta \right) =a_{0}\eta +a_{1}\eta ^{3}+a_{2}\eta
^{5}+\ldots  .
\end{equation}

Similarly, we can obtain series approximate solution for $A\neq 0$.
Here we obtain that $\alpha =0$ and $Q(0)=A=a_0, Q'(0)=B=a_1$ for
$f(\eta )=Q(\eta ^\beta )$. The differential equation (\ref{vegs})
yields
\begin{equation}
K\beta \eta ^{\gamma }\left[ \left( \beta -1\right) Q^{\prime
}+\beta \eta
^{\beta }Q^{\prime \prime }\right] \beta ^{\frac{r}{2}}Q^{\prime \frac{r}{2}%
}+\frac{1-r}{2}\beta \eta ^{\beta }Q^{\prime }+\frac{1+r}{2}Q=0,
\end{equation}
where  $\gamma =\beta -2+\left( \beta -1\right) \frac{r}{2}$. The
system of Briot-Bouquet equations  is formulated as follows
\begin{equation}
\left.
\begin{array}{c}
u_{1}(\xi ,z_{1}(\xi ),z_{2}(\xi ))=\xi \,z_{1}^{\prime }(\xi )
\\
u_{2}(\xi ,z_{1}(\xi ),z_{2}(\xi ))=\xi \,z_{2}^{\prime }(\xi )
\end{array}%
\right\}
\end{equation}%
where
\begin{equation}
\xi z^{\prime \prime }\left( \xi \right) =\frac{1-\beta }{\beta }Q^{\prime }-%
\frac{1-r}{2K}\xi ^{1-\frac{\gamma }{\beta }}\frac{Q^{\prime ^{1-\frac{r}{2}%
}}}{\beta ^{1+\frac{r}{2}}}-\frac{1+r}{2K}\frac{Q}{\beta ^{2+\frac{r}{2}%
}Q^{\prime \frac{r}{2}}}\xi ^{-\frac{\gamma }{\beta }}.
\end{equation}

The conditions in the Briot-Bouquet theorem can be satisfied for
arbitrary $A$ and $B$ if $\gamma =-\beta ,$ i.e., $\beta =1.$ Then
we get that
\begin{equation}
u_{2}(\xi ,z_{1}(\xi ),z_{2}(\xi ))=-\frac{1-r}{2K}\xi ^{2}(a_{1}+z_{2})^{1-%
\frac{r}{2}}-\frac{1+r}{2K}\xi \frac{a_{0}+a_{1}\xi +z_{1}}{(a_{1}+z_{2})^{%
\frac{r}{2}}}.
\end{equation}
The method of calculation of the coefficients in the power series is
similar to case $A=0.$ Equating the like powers one gets the further
coefficients that
\begin{equation}
a_{2}=-\frac{1+r}{4K}\frac{a_{0}}{a_{1}^{\frac{r}{2}}}, \ \ a_{3}=-\frac{\left( 1+r\right) ^{2}r}{48K^{2}}\frac{a_{0}^{2}}{a_{1}^{r}}-%
\frac{a_{1}^{1-\frac{r}{2}}}{6K}, \ \ \ldots . \end{equation}
Figure 2 presents the series solution of (11) for different values of $r$ when $A=0$ and  $B=1$. 
 Figure 3 presents the series solution of (11) for different values of $r$ when $A=1$ and  $B=0$.
Note, the red curve on both figures presents the solution for $r=-0.425$, which was presented in the phase plane analysis as well. 
Unfortunatelly, our phase plane and the series expansion results cannot be compared directly. 
\begin{figure}
\hspace*{2.4cm}
\scalebox{0.8}{
\rotatebox{0}{\includegraphics{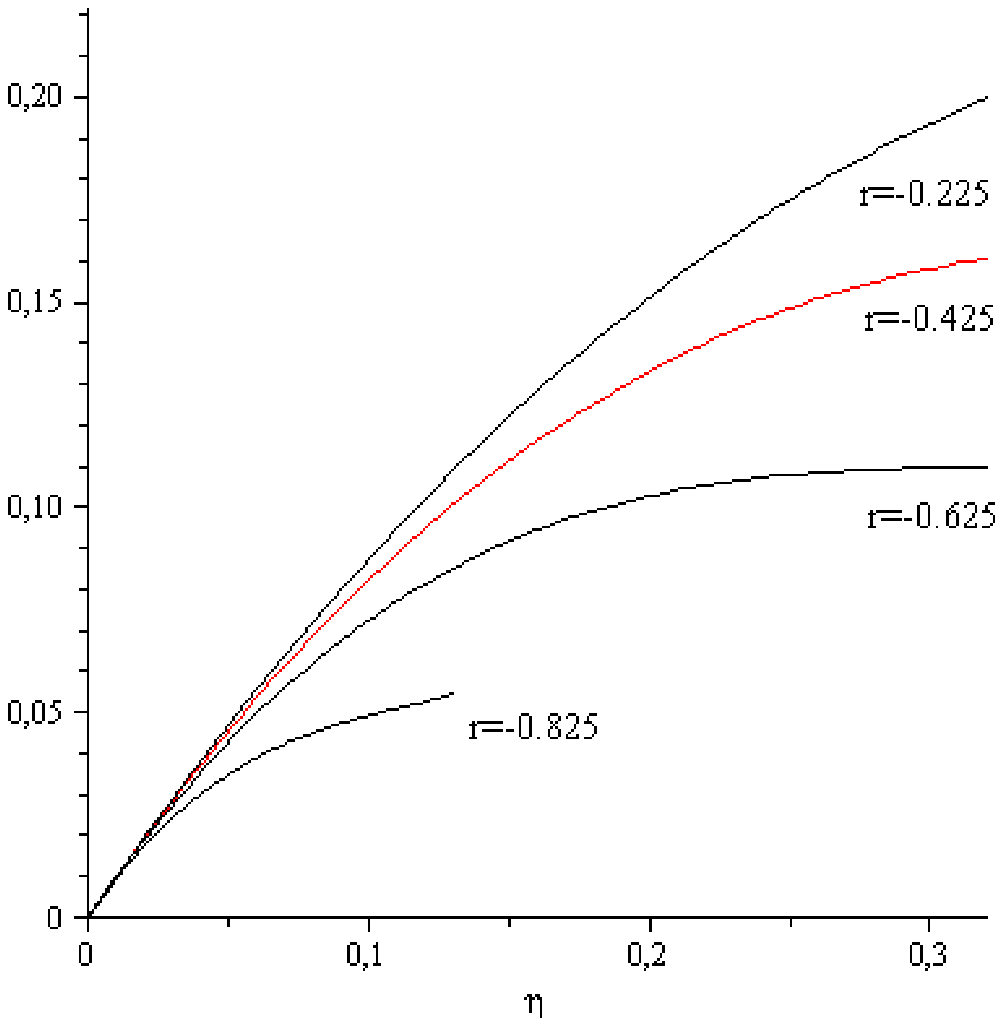}}}
\vspace*{-0.4cm}
\caption{Approximate series solutions of $f(\eta)$  to Eq. (11) and for different values of $r$  when  $A=0$  and   $B=1$. The red curve is for $r = -0.425$. } 
\label{kettes}       
\end{figure}

\begin{figure}
\hspace*{2.4cm}
\scalebox{0.8}{
\rotatebox{0}{\includegraphics{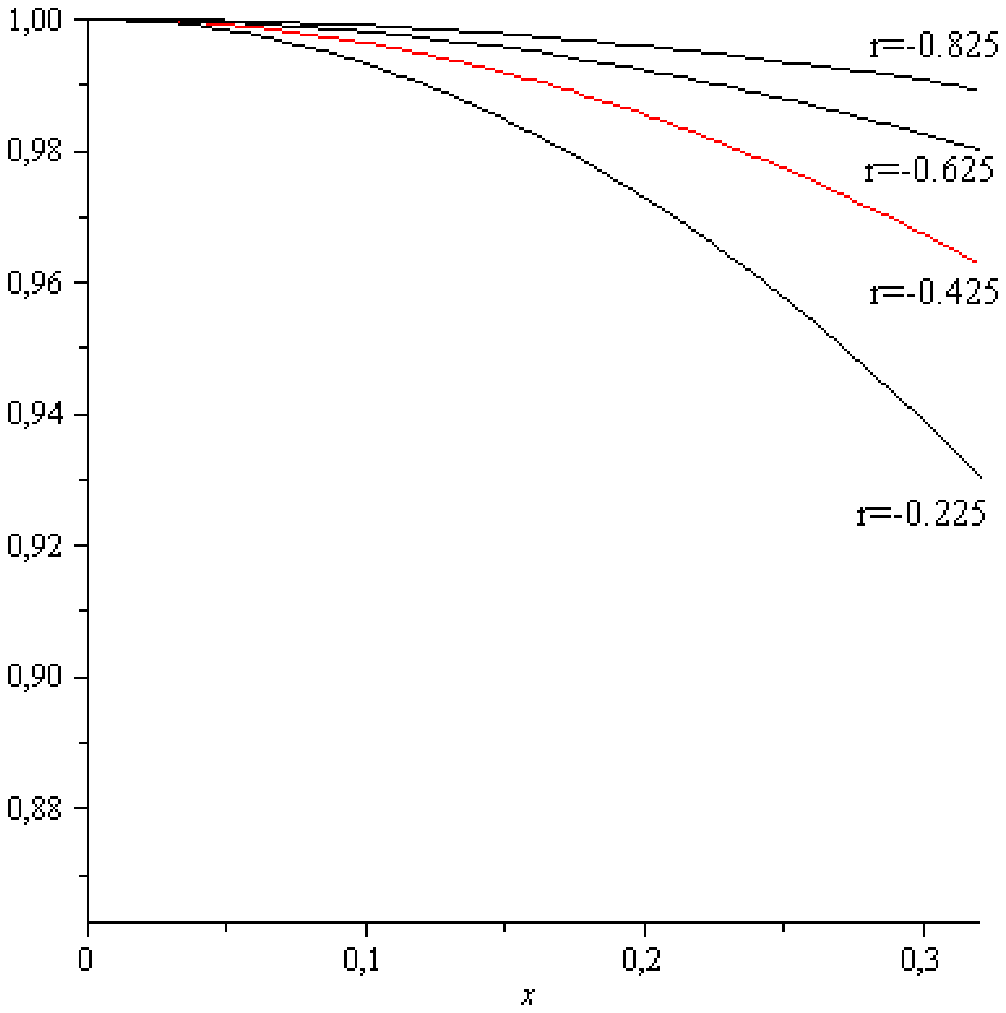}}}
\vspace*{-0.4cm}
\caption{ Approximate series solutions of $f(\eta)$ to Eq. (11) and for different 
values of $r$  when  $A=1$  and   $B=0$. Note the red curve is for $r = -0.425$.  } 
\label{kettes}       
\end{figure}

\section{Conclusions} 
We applied a two-dimensional generalization of the self-similar Anstaz for the Ladyzhenskaya-type non-Newtonian NS equation. We analyzed the final highly non-linear ODE in the phase plane and gave an approximate series expansion solutions as well. \\ 
Our main result is that the velocity field of the fluid - in contrast to our former Newtonian result - 
has a compact support, which is the major difference. \\
 We can explain it with the following everyday example:  
 Let's consider two pots in the kitchen, one is filled with water and the other is filled with chocolate cream. 
Start to stir both with a wooden spoon in the middle, after a while the whole mass of the
water becomes to move due to the Newtonian viscosity, however the chocolate cream far from the spoon remains stopped even after a long time.   	 

\section{Acknowledgement}
 This work was supported by the Hungarian OTKA NK 101438 Grant. This research was (partially) carried out in the framework of the Center of Excellence of Innovative Engineering Design and Technologies at the University of Miskolc. 
We thank for Prof. Robert Kersner for useful discussions and comments. 
We dedicate this paper to our spouses. 

\end{document}